\documentclass[10pt,a4paper]{article}
\usepackage{amsmath}
\usepackage{amsfonts}
\usepackage{amssymb}
\usepackage[dvips]{graphicx}
\usepackage{bm}

\begin{document}

\title{\textbf{Semi-classical black holes with large $N$ re-scaling and information loss problem}}
\author{\textsc{Dong-han Yeom}$^{a,b,c}$\footnote{innocent.yeom@gmail.com} and \textsc{Heeseung Zoe}$^{a,d,e}$\footnote{heezoe@gmail.com}\\
\textit{$^{a}$\small{Department of Physics, KAIST, Daejeon 305-701, Republic of Korea}}\\
\textit{$^{b}$\small{Center for Quantum Spacetime, Sogang University, Seoul 121-742, Republic of Korea}}\\
\textit{$^{c}$\small{Research Institute for Basic Science, Sogang University, Seoul 121-742, Republic of Korea}}\\
\textit{$^{d}$\small{Department of Physics, Izmir Institute of Technology, 35437 Izmir, Turkey}}\\
\textit{$^{e}$\small{Department of Physics, Middle East Technical University, 06531 Ankara, Turkey}}}
\maketitle

\begin{abstract}
We consider semi-classical black holes and related re-scalings with $N$ massless fields.
For a given semi-classical solution of an $N=1$ universe, we can find other solution of a large $N$ universe by the re-scaling.
After the re-scaling, any curvature quantity takes a sufficiently small value without changing its causal structure.
Via the re-scaling, we argue that black hole complementarity for semi-classical black holes cannot provide a fundamental resolution of the information loss problem, and the violation of black hole complementarity requires sufficiently reasonable amounts of $N$.
Such $N$ might be realized from some string inspired models.
Finally, we claim that any fundamental resolution of the information loss problem should resolve the problem of the singularity.
\end{abstract}

\newpage

\tableofcontents

\newpage

\section{Introduction}

The information loss problem of black holes was motivated by semi-classical calculations \cite{Hawking:1976ra}.
Applying quantum field theory in a classical metric of a black hole, it is observed that the black hole is evaporating by emitting Hawking radiation \cite{Hawking:1974sw}.
This calculation poses a very profound question to the unitarity of quantum mechanics.
Even though we do not have the final answer due to the absence of quantum gravity, we can advance the problem by constructing and speculating upon different semi-classical black hole solutions.

Black hole complementarity is a typical example of such reasoning \cite{Susskind:1993if}.
It reflects the non-locality which quantum gravity may contain in a certain form \cite{Lowe:1995ac} and provides rich implications to be considered in constructing quantum gravity.
However, its motivation is essentially semi-classical.
According to black hole complementarity, after the information retention time \cite{Page:1993wv}\cite{Page:1993df} (when the initial area of a black hole decreases to its half value), an observer outside of the black hole can see the information of the in-falling matter via the Hawking radiation.
However, since the free-falling information is not affected by the Hawking radiation, two copies of information may exist; this appears to violate the no cloning theorem of quantum mechanics \cite{Susskind:1993mu}.
Black hole complementarity argues that this does not pose a problem, since these two copies cannot be observed by a single observer.
In a Schwarzschild black hole, this assertion appears to hold well \cite{Susskind:1993mu}.

We introduce an interesting semi-classical setup to discuss the information loss problem: \textit{semi-classical gravity with a large number of massless fields.}
This kind of setup has already been considered to clarify semi-classical approximations for quantization of gravity.
If there is a sufficiently large number of massless fields, they will be dominant over the gravitons, and can thereby justify semi-classical approaches \cite{wald}.
However, in the context of whole quantum gravity, it is still questionable whether it is reasonable to use the limit of many massless fields.
String theory will give a low energy effective action, which contains the gravity as well as some matter fields, and the action will allow semi-classical gravity in general.
If string theory can allow a sufficiently large number of massless fields, then the large number setup will be helpful to understand the information loss problem \cite{Yeom:2008qw}, since any fundamental resolution of the information loss problem should be valid even in that extreme case.

We introduce re-scalings of semi-classical solutions between different universes: for a given semi-classical solution of an $N=1$ universe, we find another solution of a large $N$ universe by using re-scaling.
Here, after re-scaling, any curvature quantity takes a sufficiently small value without changing its causal structure.
In this context, we argue that black hole complementarity and violation of locality fail to provide a fundamental resolution of the information loss problem.

This paper is organized as follows.
In Section~\ref{sec:ashort}, we introduce the information loss problem and concepts of the information retention time, a duplication experiment, black hole complementarity, and the scrambling time.
In Section~\ref{methods}, we introduce the re-scaling and its consequences and applications.
In Section~\ref{app}, we discuss some applications of the re-scaling to the information loss problem and
in Section~\ref{dis}, we conclude that the information loss problem inevitably related to the problem of the singularity.

\section{\label{sec:ashort}A short introduction to the information loss problem}

In this section, we discuss some preliminaries to understand the information loss problem of black holes.

\subsection{The information loss problem of a black hole}

The information loss problem of a black hole was initiated by the developments of general relativity \cite{wald}.
According to the no-hair theorem \cite{Israel:1967wq}, which already been known, the only stationary, axisymmetric, and electrovac solution is the charged Kerr solution \cite{wald}.
The solution is determined by only three quantities: mass $M$, charge $Q$, and angular momentum $J$ of the black hole.
According to Bardeen, Carter, and Hawking \cite{Bardeen:1973gs}, black holes with $M$, $Q$, and $J$ should be related by the following equation:
\begin{eqnarray}
\delta M = \frac{\kappa}{8\pi} \delta A + \Omega \delta J,
\end{eqnarray}
where $\kappa$ is the surface gravity, $A$ is the area of the event horizon of the black hole, $\Omega$ is the angular velocity of the horizon, and $\delta$ is a variation for each functions.
Also, Hawking proved that the area of the event horizon $A$ should increase and never decrease \cite{Hawking:1971vc}:
\begin{eqnarray}
\delta A \geq 0.
\end{eqnarray}
Therefore, one may guess that a black hole can be described as a thermal system that is determined by only three quantities: $M$, $Q$, and $J$.
However, in this thermal interpretation, one missing link was the meaning of entropy, and Bekenstein guessed that the area should proportional to the entropy of a black hole \cite{Bekenstein:1973ur}.

In $1974$, Hawking calculated the number distribution of thermal radiation, which is generated by quantum effects from a black hole \cite{Hawking:1974rv}\cite{Hawking:1974sw}.
For a massless scalar field case, the particle number distribution $n_{\omega}$, which is a function of particle energy $\omega$, is proportional to the Planck distribution:
\begin{eqnarray}
\langle n_{\omega} \rangle \propto \frac{1}{\exp{\frac{2 \pi \omega}{\kappa}} - 1}.
\end{eqnarray}
Therefore, the temperature of thermal radiation is $T = \kappa/2\pi$ and the \textit{thermal} entropy should be proportional to the area $A$ with a numerical constant $1/4$:
\begin{eqnarray} \label{entropyformula}
S = \frac{A}{4}.
\end{eqnarray}

After obtaining this conclusion, Hawking guessed at the existence of the information loss problem \cite{Hawking:1976ra}. In almost all natural cases, black holes have a singularity inside them \cite{Hawking:1969sw}\cite{Hawking:1973uf}. Although it is not possible to understand beyond the singularity by general relativity, this is not so serious of a problem since we cannot see the inside of the black hole. If a black hole is forever, although we do not know inside of the black hole, one may guess that information is still inside of the black hole. However, if the black hole disappears, information inside of the event horizon, which is causally disconnected with the outside, will disappear. According to Hawking's calculation, Hawking radiation is totally thermal, and we can restore only three pieces of information from Hawking radiation: $M$, $Q$, and $J$. The other quantum information seems to disappear as a black hole evaporates.

Then, what is going on? If information disappears and only $M$, $Q$, and $J$ remain, then we cannot reconstruct the initial state from the final state. Then, the time evolution of the state cannot be described by a unitary transformation. Then, maybe quantum gravity or the fundamental theory cannot be a unitary theory, and then we will lose fundamental predictability \cite{Hawking:1976ra}.

Therefore, many physicists regarded the problem as a serious and fundamental problem to approach the Theory of Everything or quantum gravity. So far, there has been no common consensus about the problem, but there has been a great deal of progress and a number of clues to understand the problem. In the next section, we will discuss such important clues.

\subsection{Related problems}

In this section, we discuss on entropy of a black hole and an information emission from a black hole.

\subsubsection{Entropy of a black hole}

Bekenstein and Hawking's entropy is just thermal entropy. They first calculated the temperature and second defined the entropy $dS_{\mathrm{th}} = dQ/T$, where $dQ$ is the difference of heat. The natural question is then whether it is not only the thermal entropy but also the \textit{statistical} entropy, $S_{\mathrm{st}} = \log {N}$, where $N$ is the number of accessible states.

There was important progress on the issue in quantum gravity.

First, in string theory, it is known that open strings can be attached to a boundary, and the boundaries can be regarded as physical objects, so called D-branes \cite{Polchinski:1998rq}. If the coupling constant is sufficiently small, the D-brane looks like a membrane. However, if the coupling constant is sufficiently large, it becomes a gravitational object and will form a black hole type geometry \cite{Callan:1996dv}. It is known that for certain extreme limits with supersymmetry, the entropy of the weak coupling limit is the same as that of the strong coupling limit. Researchers found some combinations of D-branes that gives black hole solutions and could calculate the entropy of the weak coupling limit \cite{Strominger:1996sh}. The entropy could be exactly matched to the entropy formula in Equation~(\ref{entropyformula}) for some extreme cases.

Second, in loop quantum gravity, researchers could calculate the quantization of an area \cite{Rovelli:1994ge}. If one wants to calculate the accessible degrees of freedom of a black hole for the outside observer, it should be related to the degree of freedom of the locally defined outer horizon (the isolated horizon) \cite{Ashtekar:2004cn}. According to loop quantum gravity, the quantized area of the horizon is described by the spin network \cite{Rovelli:1989za} and the spin network allows the calculation of the number of states for the area. Consequently, it is not so strange that the statistical entropy is proportional to the area. The proportional constant $1/4$ can be obtained \cite{Rovelli:1996dv} if we carefully choose a constant (the Immirzi parameter). Although we do not know why we have to choose such a parameter, this approach has a strong point since loop quantum gravity could calculate the statistical entropy of Schwarzschild black holes, while string theory was only usable for calculation of certain extreme black holes.

Third, if one believes the Euclidean analytic continuation is sound, one can calculate the entropy of a black hole in a simple way by the Euclidean signatures. The coordinate time is not well-defined inside of the horizon, and, hence, it is convenient to choose a coordinate in which the coordinate begins at the horizon. With this treatment, we have to add a boundary term (the Gibbons-Hawking boundary term) in the action as a price for the choice of the `bad' coordinate. Then, the Euclidean space-time and the action integral are well-defined and the action integral corresponds to the entropy of the black hole \cite{Gibbons:1976ue}. Although the entropy in this approach resembles the thermal entropy, if we trust the analytic continuation, it will yield the correct result as a non-perturbative quantum gravitational approach.

Therefore, although there is no formal proof on the thermal and statistical entropy relation,
\begin{eqnarray}
\frac{A}{4} = \log N,
\end{eqnarray}
it is quite natural to believe the correspondence. In the following subsections, we will assume that the area of a black hole is proportional to the statistical entropy and study consequences of the assumption.

\subsubsection{Information emission from a black hole}

The statistical entropy is the capacity of information in terms of an information theoretical sense. Therefore, one can guess that if the statistical entropy decreases in a subsystem, then the subsystem should emit information at a certain time before it lose information.

To represent this intuition, we need to define information that is conserved by any unitary processes. Lloyd and Pagels \cite{Lloyd:1988cn} defined information $I$ by the following formula:
\begin{eqnarray}
I = S^{(\mathrm{course-grained})} - S^{(\mathrm{fine-grained})},
\end{eqnarray}
where $S^{(\mathrm{course-grained})}$ is the course-grained entropy and $S^{(\mathrm{fine-grained})}$ is the fine-grained entropy. Its meaning is intuitively clear. If I have $5$ MB of memory, $5$ MB is the maximum capacity of information, and it is the course-grained entropy. However, if the memory already contains a $3$ MB file, I can use only $2$ MB in real situations and this really accessible information capacity is the fine-grained entropy. Then, the difference between the course-grained entropy and the fine-grained entropy is information: $5-2=3$ MB.

Then, the natural step is to calculate the course-grained and the fine-grained entropy \cite{Page:1993wv}. Let us consider a system with a number of degrees of freedom $m \times n$ and divide by two subsystems, $A$ and $B$, where $A$ has a number of degrees of freedom $m$ and $B$ has $n$. For subsystem $A$, its maximal entropy is $\log m$ and it is the course-grained entropy of $A$. However, because of the entanglement between $A$ and $B$, one cannot use all of $m$ and it is limited. To calculate the really accessible entropy, first we have to trace out the degrees of freedom of $B$, and second we calculate the entropy of the subsystem:
\begin{eqnarray}
\rho_{A,B} &\equiv& \mathrm{tr}_{B,A} \rho,\\
S_{A,B}^{(\mathrm{fine-grained})} &=& - \mathrm{tr} \rho_{A,B} \log \rho_{A,B},
\end{eqnarray}
where $\rho$ is the density matrix of the total system. It is easy to prove that $S_{A}^{(\mathrm{fine-grained})} = S_{B}^{(\mathrm{fine-grained})}$, if the total system is in a pure state.

Let us consider that $m$ and $n$ vary with time and initially $n$ was zero. Then, the total initial information was contained by $A$. Now, as the degree of freedom of $A$ decreases and it is emitted to $B$, $B$ feels that $A$ emits information by the formula:
\begin{eqnarray}
I_{B} = \log n - \left(- \mathrm{tr} \rho_{B} \log \rho_{B}\right).
\end{eqnarray}
Then, as $A$ emits particles, $m$ decreases, $n$ increases, and, hence, information emitted from $A$ to $B$ increases.

\begin{figure}
\begin{center}
\includegraphics[scale=0.8]{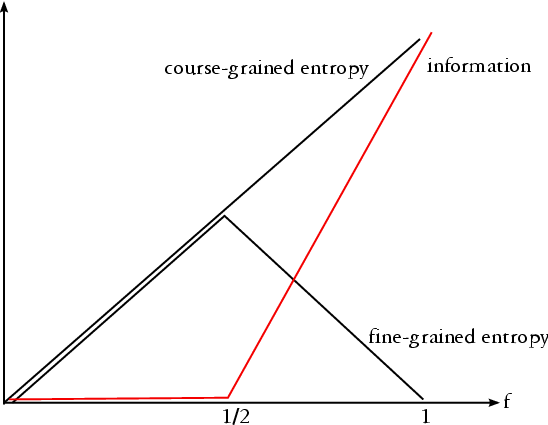}
\caption{\label{fig:information_retention}Emission of information, where $f$ is the fraction of the escaped course-grained entropy to the original course-grained entropy.}
\end{center}
\end{figure}

We can estimate further if we assume that the total system is pure and random. Page conjectured the following \cite{Page:1993df} and soon after it was proven \cite{Sen:1996ph}: if $1 \ll n \leq m$, then
\begin{eqnarray}
S_{A}^{(\mathrm{fine-grained})} &=& \sum_{k=m+1}^{mn} \frac{1}{k} - \frac{n-1}{2m}\\
&\cong& \log n - \frac{n}{2m}.
\end{eqnarray}
Therefore, initially, information is emitted $\cong n/2m$, and it is negligible.
If $n > m$, since $S_{A}^{(\mathrm{fine-grained})} = S_{B}^{(\mathrm{fine-grained})}$,
\begin{eqnarray}
S_{A}^{(\mathrm{fine-grained})} &=& \sum_{k=n+1}^{mn} \frac{1}{k} - \frac{m-1}{2n}\\
&\cong& \log m - \frac{m}{2n}.
\end{eqnarray}
Therefore, after $n$ becomes greater than $m$, the emitted information is $\cong \log n - \log m + m/2n$, and it gradually increases (Figure~\ref{fig:information_retention}).

This result is quite strong since our assumption is only for pure and random states. As a conclusion, a system $A$ begins to emit information to $B$ when its course-grained entropy decreases its half value ($m=n$). Before that time, emitted particles do not have sufficient information. However, after that time, the original information cannot be compressed to $A$ and the information of $A$ has to be transferred to $B$ by the emitted particles.

\subsubsection{Conclusion}

If we assume the results of the previous two subsections so that $A/4 = \log N$ and a black hole begins to emit information when $\log N \rightarrow (1/2) \log N$, then we can conclude that the black hole begins to emit information when its area decreases to the half of the initial area.

The time scale is the order of the lifetime of a black hole $\sim M^{3}$: this time is called the \textit{information retention time} \cite{Susskind:1993mu}. This time scale is sufficiently large, but in many cases, the black hole can be still semi-classical, i.e., even though the area of the black hole decreased to half its value, the black hole is still large enough. Then, the only way to take out information from the large black hole is Hawking radiation. Therefore, information should be emitted by Hawking radiation if we assume the previous two subsections.

\subsection{Black hole complementarity}

If we assume and accept the discussions of the previous section, we obtain that information should be contained by Hawking radiation. Also, let us further assume that there is an observer such that the observer can figure out information from the Hawking radiation. Then, the information should be emitted and the outside observer can detect information.

\begin{figure}
\begin{center}
\includegraphics[scale=0.8]{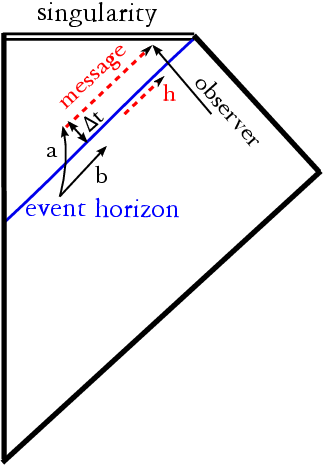}
\caption{\label{fig:Schwarzschild_duplication}The duplication experiment. $a$ and $b$ are a pair of correlated spins. The observer sees $h$, which is a copy of $a$ after the information retention time via Hawking radiation. To see $a$, $a$ should be sent to the out-going direction after the time $\Delta t$. If the observer sees both $a$ and $h$, since they are both correlated to $b$, it violates the no-cloning theorem and unitarity.}
\end{center}
\end{figure}

Let us think of a specific situation (Figure~\ref{fig:Schwarzschild_duplication}) \cite{Susskind:1993mu}. Let us consider a series of experiments in which a pair of correlated spins are created outside of the event horizon. Let us call that one of the pair that falls into the black hole $a$ and the other of the pair that is outside of the black hole $b$. If Hawking radiation contains information, then information about $a$ can be emitted by Hawking radiation, and we call it $h$. If there is an observer who can measure the state of $h$, falls into the black hole, and measures the state of $a$, then eventually we will know that the collected information $a$ and $h$ are both correlated to $b$. This implies that the observer sees the duplication of states, which is disallowed by quantum mechanics. We will call this kind of experiment as a \textit{duplication experiment}.

Susskind and Thorlacius \cite{Susskind:1993mu} could answer questions on the duplication experiment. If the observer sees both $a$ and $h$, the observer has to wait until the information retention time. However, if the original free-falling information $a$ touches the singularity of the black hole, then there is no hope to see the duplication. To see the duplication, the free-falling information $a$ should be sent to the out-going direction during the time $\Delta t$.

We can estimate the time $\Delta t$ in the Schwarzschild space-time. In general, $(4+n)$-dimensional Schwarzschild black holes are described by the following metric \cite{Lowe:1999pk}:
\begin{eqnarray}
ds^{2} = - \left( 1- \frac{\mu}{r^{2}}r^{1-n} \right) dt^{2} + \frac{1}{\left( 1- \frac{\mu}{r^{2}}r^{1-n} \right)}
dr^{2} + r^{2} d\Omega^{2},
\end{eqnarray}
where $\mu$ is a parameter that is related to mass and $\Omega$ is a solid angle of an $(n+2)$-sphere.
The horizon is $r_{0} = \mu^{1/(n+1)}$, and it is not difficult to confirm that the Hawking temperature is on the order of $T \sim 1/r_{0}$. Therefore, if one considers a lifetime of a $(4+n)$-dimensional black hole, one can easily calculate that
\begin{eqnarray}
\frac{dM}{dt} \sim \frac{d \mu}{dt} \sim A_{(4+n)} T^{(4+n)} \sim r_{0}^{2+n} \frac{1}{r_{0}^{4+n}} \sim
\frac{1}{r_{0}^{2}},
\end{eqnarray}
and one obtains a lifetime $\tau \sim \mu^{\frac{n+3}{n+1}} \sim r_{0}^{n+3}$.

For the next calculation, we will comment on a simple extension to Kruskal-Szekeres coordinates \cite{Susskind:1993mu}\cite{Lowe:2006xm}. Let us assume that we neglect the angular part, and we assume the form
\begin{eqnarray}
ds^{2} = F(R) \left( -R^{2} d\omega^{2} + dR^{2} \right).
\end{eqnarray}
To compare the original metric, the following assumptions are reasonable:
\begin{eqnarray}
d \omega^{2} &=& \frac{dt^{2}}{r_{0}^{2}}, \nonumber \\
R^{2}F(R) &=& r_{0}^{2} \left( 1- \frac{\mu}{r^{2}}r^{1-n} \right), \nonumber \\
F(R)dR^{2} &=& \frac{1}{\left( 1- \frac{\mu}{r^{2}}r^{1-n} \right)} dr^{2}.
\end{eqnarray}
In terms of the coordinate $R$, the singularity occurs at $R^{2} = - r_{0}^{2}$; and the horizon occurs at $R=0$.

Now, we can choose another metric and coordinates $(U,V)$ by
\begin{eqnarray}
V &=& R e^{\omega}, \nonumber \\
U &=& - R e^{-\omega}, \nonumber \\
ds^{2} &=& -F(R) dU dV.
\end{eqnarray}
Here, the singularity is $UV = r_{0}^{2}$.

Now, we can state the condition of a duplication experiment in a Schwarzschild black hole \cite{Susskind:1993mu}.
The first observer falls into a black hole and sends a signal to the out-going direction around time $\Delta t$.
Now assume that a second observer hovers above the horizon at a distance of the order of the Planck length $l_{\mathrm{Pl}}$
and jumps into the black hole at the information retention time $\sim \tau$.
Then, the initial location of the second observer is $V=R e^{\omega}$,
where $R \sim l_{\mathrm{Pl}}$ and $\omega \sim \tau / r_{0}$.
Before touching the singularity, the second observer will spend time (in terms of $U$) around $\sim r_{0}^{2} / V$
since the singularity is $UV = r_{0}^{2}$.
Therefore, the first observer should send a signal around time $\Delta t \sim e^{- \tau / r_{0}}$.

Therefore, the duplication may be observed if one can send a signal between the time
\begin{eqnarray}
\Delta t \sim \exp{-\frac{\tau}{r_{0}}} \sim \exp{-\frac{\tau}{M}},
\end{eqnarray}
where $\tau$ is the information retention time ($\sim M^{3}$ for $4$-dimensional cases).

Then, to send a quantum bit during $\Delta t$, it has to satisfy the uncertainty relation $\Delta t \Delta E \gtrsim 1$. Then, for $4$-dimensional cases, the required energy to send a quantum bit of information during $\Delta t$ is $\sim \exp M^{2}$, which is greater than the original mass of the black hole $M$. Therefore, the duplication experiment seems to be impossible in real situations \cite{Susskind:1993mu}.

If we accept the results of Susskind and Thorlacius, the idea that Hawking radiation contains information is self-consistent, although it seems to duplicate information in the inside and the outside of the black hole. Although information is duplicated, if no observer can see the violation of the natural laws, there is no problem. In other words, there is no global description for both a free-falling observer and an asymptotic observer, and we have to choose one of them. This seems to be a contradiction, but there is effectively no problem since no one can observe both situations. Therefore, in this sense, two observers are complementary. This principle is known by \textit{black hole complementarity} or observer complementarity \cite{Susskind:1993if}.

Black hole complementarity is consistent with two paradigms: the membrane paradigm \cite{Thorne:1986iy} and the D-brane picture \cite{Callan:1996dv}. The membrane paradigm is to see a black hole as a membrane around the event horizon, the so-called stretched horizon. If we send an object to a black hole, the object is stretched and scrambled on the horizon. The outside observer cannot see the object disappear on the horizon. Therefore, for the outside observer, information is on the horizon and eventually escapes from the black hole via Hawking radiation. Here, the scrambling occurs in the following order of time:
\begin{eqnarray}
\tau_{\mathrm{scr}} \sim M \log M,
\end{eqnarray}
and this time is called the \textit{scrambling time} \cite{Sekino:2008he}. According to Hayden and Preskill \cite{Hayden:2007cs}, after a black hole approaches the information retention time, if one sends small bits of information, this information will quickly escape the black hole after the scrambling time. The D-brane picture is to see a black hole as a combination of D-branes in the strong coupling limit. Then, one can see non-trivial correspondences between normal black holes and D-branes \cite{Callan:1996dv}. However, such correspondences require special symmetry and, hence, are limited for extreme or near-extreme charged black holes.

According to the black hole complementarity principle, for the outside observer, information is attached to the stretched horizon, and the horizon thermalizes and emits information. For the inside observer, information freely falls and touches the singularity. If we choose one observer, then they can be described by general relativity and local quantum field theory, and hence in terms of observations, there is no contradiction on general relativity and local quantum field theory \cite{Susskind:1993if}. However, if one wants to describe both of inside and outside, or if one wants to understand how to reconcile both of inside and outside, one has to study the correlation between the inside and the outside \cite{Lowe:1995ac}. The relation should be \textit{non-local} for not only small scales but also large scales \cite{Giddings:2001pt}. (Here, the term `non-local' means that there is a correlation function that is not vanishing for a space-like separation.) We do not have consensus on the non-local correlations, but quantum entanglement \cite{Horowitz:2003he}, string theory \cite{Susskind:1993if}\cite{Iizuka:2008hg}, or somehow strong gravity regimes \cite{Giddings:2001pt} of a certain type of gravity theories can be candidates for the non-local correlations.

\section{\label{methods}Methods of large $N$ re-scaling}

\subsection{A simple model}

Let us think a Schwarzschild black hole in $4$-dimensions:
\begin{eqnarray}
ds^{2} = - \left( 1- \frac{2M}{r} \right) dt^{2} + \left( 1- \frac{2M}{r} \right)^{-1} dr^{2} + r^{2} d\Omega^{2}.
\end{eqnarray}
Then, the size of the black hole is $2M$ and the lifetime of the black hole is order of $M^{3}$.

First, let us observe a heuristic example. (1) If there is a black hole with mass $M=100$ in the Planck unit and we have only $N=1$ field that contributes to Hawking radiation, then its radius is $2M = 200$ and its lifetime is $M^{3}/N=1000000$ (we ignore the common constant factor). (2) If there is a black hole with mass $100$ in the Planck unit and we have $100$ fields, then its radius is $200$ and its lifetime is proportional to $10000$. (3) If there is a black hole with mass $1000$ in the Planck unit and we have $100$ fields, then its radius is $2000$ and its lifetime is proportional to $10000000$.

If we compare Case (1) and Case (3), they are physically different (masses in the Planck unit and number of fields in each universes). However, in terms of causal structures, they are similar because their ratios between the temporal size (lifetime $\tau$) and the spatial size (radius $r_{0}$) are the same: $\tau/r_{0} = 1000000/200 = 10000000/2000$.

Now, what is the difference between Case (2) and Case (3)? The number of fields are the same. The difference is a constant factor on the mass. The question is that, when we change the number of fields, what should be the factor of the mass to maintain the ratio between the temporal size and the spatial size of a black hole? The answer is $\sqrt{N}$ for $4$-dimensions.

\begin{figure}
\begin{center}
\includegraphics[scale=0.7]{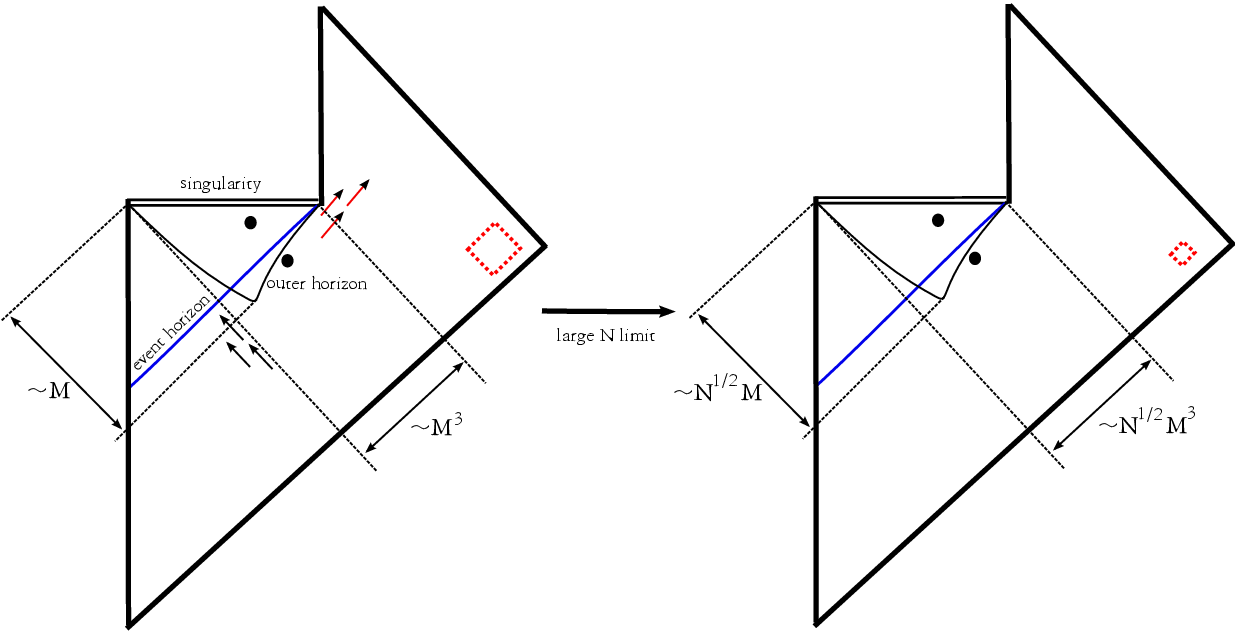}
\caption{\label{fig:Sch} The causal structure of Schwarzschild black holes for the mass $M$ with one field Hawking radiation case and the mass $\sqrt{N}M$ with $N$ field Hawking radiation case. The spatial size and the temporal size are stretched by factor $\sqrt{N}$. The red squares are equal length squares for the one field Hawking radiation case and the four field Hawking radiation case. For large $N$ case, the equal length squares becomes relatively smaller and smaller. The black dots each diagram are conformaly equivalent between the $N=1$ case and the large $N$ case. Therefore, the distance between conformaly equivalent points should be stretched by the $\sqrt{N}$ factor.}
\end{center}
\end{figure}

Let us restate this phenomenon. Let us imagine a situation that there is a number $N$ of fields which are contribute to Hawking radiation. If we assume a black hole with mass $M$, then its lifetime is order of $M^{3}/N$ via $N$ fields. Then what will happen if we think a black hole with mass $\sqrt{N}M$ with $N$ fields? The size of the black hole is $2M\sqrt{N}$ in the Planck unit. The lifetime of the black hole is order of $(\sqrt{N}M)^{3}/N = \sqrt{N}M^{3}$ in the Planck unit. Therefore, the spatial size and the temporal size are \textit{both} stretched $\sqrt{N}$ times more than these of the mass $M$ and $N=1$ field case (Figure~\ref{fig:Sch}).

Therefore, one can guess that the causal structure of the large $N$ limit of the mass $\sqrt{N}M$ black hole and the causal structure of the mass $M$ black hole with one field Hawking radiation case are the same, although the former is relatively larger in the Planck units than the latter. It is not inconsistent: if we stretch the length parameter $r$ and the time parameter $t$ by the factor $\sqrt{N}$ and stretch the mass by $\sqrt{N}M$, then the metric is
\begin{eqnarray}
ds^{2} = N \left[ - \left( 1- \frac{2M}{r} \right) dt^{2} + \left( 1- \frac{2M}{r} \right)^{-1} dr^{2} + r^{2} d\Omega^{2} \right],
\end{eqnarray}
and hence it is a solution of the Einstein equation again and the causal structures of any $N$ are conformaly equivalent. Of course, it should be checked whether this relation holds for more general and dynamical situations.

For each $N$, if causal structures of $\sqrt{N}M$ black holes with $N$ field Hawking radiation are conformaly invariant, one can obtain two conclusions:
\begin{enumerate}
\item As the size and the lifetime are stretched by factor $\sqrt{N}$, any distances between conformaly equivalent points are stretched together by factor $\sqrt{N}$;
\item As the size and the lifetime are stretched by factor $\sqrt{N}$, the curvatures (e.g., the Kretschmann scalar $R_{abcd}R^{abcd}$) on conformaly equivalent points will be smaller and smaller by factor $1/N^{\alpha/2}$, where $\alpha$ is the length dimension of the inverse of the curvature (for the Kretschmann scalar, $\alpha=4$; for the Ricci scalar, $\alpha=2$), since the space and time are stretched by factor $\sqrt{N}$.
\end{enumerate}

In the following subsections, we will justify these intuitions to more formal and general situations of semi-classical theory.

\subsection{The semi-classical theory}

Let us assume that a low energy effective action is given as follows:
\begin{eqnarray} \label{condition1}
S = (\textrm{4-D gravity}) + (\textrm{a large number of massless fields}).
\end{eqnarray}
For simplicity, let us assume that we make a black hole with one massless scalar field. (All of the arguments of this section can easily be applied to general classical field configurations.)
The equation of motion for one field is as follows:
\begin{eqnarray}
\phi_{;ab}g^{ab} = 0.
\end{eqnarray}
And the Einstein equation can be written as the expansion of $\hbar$:
\begin{eqnarray} \label{Einstein_largeN}
G_{\mu \nu} = 8 \pi G_{4} (T_{\mu \nu} + N \hbar \langle T^{(1)(1-loop)}_{\mu \nu} \rangle + \mathcal{O}(\hbar^{2})),
\end{eqnarray}
where $T_{\mu \nu}$ is the stress tensor for classical field configurations and $\langle T^{(1)(1-loop)}_{\mu\nu} \rangle$ is the $1$-loop order re-normalized stress tensor of \textit{one} massless field. Here, the energy-momentum tensor of the classical part becomes
\begin{eqnarray}
T_{ab}=\phi_{;a}\phi_{;b}-\frac{1}{2}\phi_{;c}\phi_{;d}g^{cd}g_{ab},
\end{eqnarray}
and $N$ is the number of massless fields.
Here, we can regard that the contribution of graviton is suppressed via sufficiently large number of matter fields \cite{wald}.

We will assume that each massless fields are independently contribute to Hawking radiation.
Then, Equation~(\ref{Einstein_largeN}) becomes reasonable for each $N$.
For example, if each scalar fields couple to gravity only and do not interact each other, the form of $N \langle T \rangle$ is justified for all loop orders. Of course, one may worry that we may not assume the independency of each fields for an extremely large $N$ case. However, for sufficiently reasonable $N$, this is still a reasonable assumption. In the following sections, we will discuss that how much we reduce the number of $N$ to a reasonable range for our purpose.

\subsection{The scheme of re-scaling}

Let us assume that $G_{4}=c=1$ and remain $\hbar$ explicitly. Then, all length, mass, and time dimensions are the same.

First, let us assume $N=1$. Then the Einstein equation becomes
\begin{eqnarray} \label{semi-classical}
G_{\mu \nu} = 8 \pi (T_{\mu \nu} + \hbar \langle T_{\mu \nu} \rangle).
\end{eqnarray}
If we call $L$ as a length dimension, we know that $[T^{ab}]=L^{-2}, [\phi_{,a}\phi^{,ab}]=L^{-3}, [R^{a}_{\;a}]=L^{-2},$ etc.

Now we define the re-scaling using the following law: if a quantity $X$ which does not explicitly depend on $\hbar$ has a dimension $[X]=L^{\alpha}$ with a certain number $\alpha$, we define a re-scaled $X'$ by
\begin{eqnarray} \label{eq:res}
X' = \sqrt{N^{\alpha}} X.
\end{eqnarray}
For example, the Ricci scalar $R$ is a quantity that has a dimension of curvature $L^{-2}$ and the re-scaled value will be $R/N$.

Then, we claim that if we re-scale \textit{all} possible quantities, then the re-scaled quantities are solutions of the following equation:
\begin{eqnarray} \label{semi-classical_2}
G'_{\mu \nu} = 8 \pi (T'_{\mu \nu} + N \hbar \langle T'_{\mu \nu} \rangle).
\end{eqnarray}
This is easy to check: $G_{\mu \nu}$ has a dimension $L^{-2}$, $T_{\mu \nu}$ has a dimension $L^{-2}$, and $\langle T_{\mu \nu} \rangle$ has a dimension $L^{-4}$ in the one-loop order. Hence, $G'_{\mu \nu} = G_{\mu \nu}/N$, $T'_{\mu \nu} = T_{\mu \nu}/N$, and $\langle T'_{\mu \nu} \rangle = \langle T_{\mu \nu} \rangle / N^{2}$. Then,
\begin{eqnarray} \label{semi-classical_3}
G_{\mu \nu} &=& N G'_{\mu \nu} \nonumber \\
&=& 8 \pi (T_{\mu \nu} + \hbar \langle T_{\mu \nu} \rangle) = 8 \pi (N T'_{\mu \nu} + \hbar N^{2} \langle T'_{\mu \nu} \rangle),
\end{eqnarray}
and our claim is proved.

In conclusion, \textit{for given quantities of solutions of Equation~(\ref{semi-classical}), the re-scaled quantities are solutions of Equation~(\ref{semi-classical_2}) with $N$ massless fields.}

\subsection{\label{invariance}Invariance of causal structures}

\begin{figure}
\begin{center}
\includegraphics[scale=0.7]{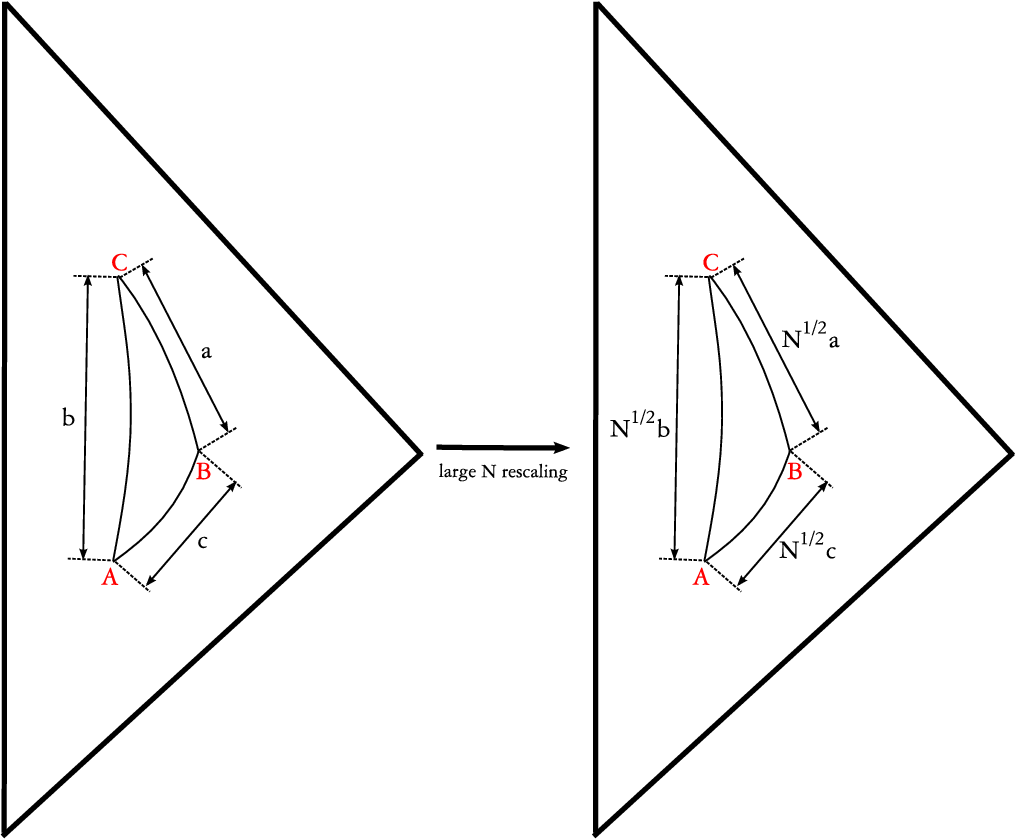}
\caption{\label{fig:invariance} Invariance of the causal structure.}
\end{center}
\end{figure}

If $ds^{2}$ is a solution of Equation~(\ref{semi-classical}), then $ds^{'2} = N ds^{2}$ is a solution of Equation~(\ref{semi-classical_2}) via the re-scaling. Let us assume that the distance between arbitrary points $A$ and $B$ along a time-like curve is $c$, $B$ and $C$ is $a$, and $A$ and $C$ is $b$ (Figure \ref{fig:invariance}). Then all of $a$, $b$, and $c$ are re-scaled to $\sqrt{N} a$, $\sqrt{N} b$, and $\sqrt{N} c$. Also, as a specific case, if one assumes a spherical symmetry (in the double null coordinate) $ds^{2} = - \alpha(u,v)^{2} dudv + r(u,v)^{2} d\Omega^{2}$, the causal structure will be determined by $r(u,v)$, and all $r(u,v)$ will be re-scaled to $\sqrt{N} r(u,v)$. Therefore, the causal structures of $ds^{2}$ and $ds^{'2}$ are invariant up to this re-scaling.

Two important remarks are noted here. First, the re-scaling conserves the causal structure of the metric. Therefore, \textit{we can use the same Penrose diagram of the $N = 1$ case.} Second, if we can prepare a sufficiently large $N$ universe, \textit{even if a region has large curvature in the $N = 1$ case, we can find a universe where the curvature is re-scaled to a sufficiently small value.} Therefore, if one can prepare an arbitrary large $N$ universe from string theory, any causal structure that is obtained from the semi-classical equation
$G_{\mu \nu} = 8 \pi (T_{\mu \nu} + \langle T_{\mu \nu} \rangle)$
is justified in string theory, since we can re-scale all curvatures to be sufficiently small.

\subsection{Applications to other theories}

Thus far, we have assumed that we construct a black hole with a massless field. However, in many cases, we can attempt to apply the same principles to other theories.
For example, if one wants to discuss a charged black hole, it is convenient to assume a complex massless scalar field $\phi$ and a Maxwell field $A_{\mu}$ by
\begin{eqnarray}
\mathcal{L} = - (\phi_{;a}+ieA_{a}\phi)g^{ab}(\overline{\phi}_{;b}-ieA_{b}\overline{\phi})-\frac{1}{8\pi}F_{ab}F^{ab},
\end{eqnarray}
where $F_{ab}=A_{b;a}-A_{a;b}$, and $e$ is the unit charge. One can easily check that, $e$ is re-scaled by $e/\sqrt{N}$.
Also, if one wants to discuss a black hole with a potential, one can also impose the same scheme, for example, to the following potential:
\begin{eqnarray}
V(\Phi)=A\Phi^{4} + B\Phi^{3} + C\Phi^{2}.
\end{eqnarray}
Here, each constants should be re-scaled by $1/N$.
In both cases, we can apply the same re-scaling schemes, but some tunings of parameters may be required.

Therefore, this re-scaling is universal for (at least) neutral black holes with a free scalar field, while the re-scaling is limited or needs further assumptions to apply for general field theories.

\section{\label{app}Applications}

\subsection{Schwarzschild black holes}

For a $4$-dimensional case, by the re-scaling, we will re-scale all length, mass, and time parameters by $\sqrt{N}$. Now, the lifetime $\tau$ for the mass $M$ and $1$ field case is re-scaled to $\tau'$ for the mass $M'=\sqrt{N}M$ and a large $N$, where
\begin{eqnarray}
\tau \sim M^{3},
\end{eqnarray}
\begin{eqnarray}
\tau' \sim \frac{M'^{3}}{N} = \frac{(\sqrt{N}M)^{3}}{N} = \sqrt{N}M^{3}.
\end{eqnarray}
This conclusion is the correct interpretation, since $\tau$ has a time dimension. Here, we have to divide the lifetime by $N$, since there are (effectively) $N$-independent fields contribute to Hawking radiation. Note that the size $r_{0}$ will be re-scaled by $r_{0}' = \sqrt{N}r_{0}$. Therefore, under the large $N$ re-scaling, the ratio between the temporal size and the spatial size is constant:
\begin{eqnarray}
\frac{\tau}{r_{0}} = \frac{\tau'}{r_{0}'}.
\end{eqnarray}

For a $5$-dimensional case, all length and time parameters will be re-scaled by $N^{1/3}$ and mass parameters will be re-scaled by $N^{2/3}$ (as given by Equation (\ref{D-dimension})). The horizon is $r_{0} \sim \sqrt{\mu}$. Therefore,
\begin{eqnarray}
\tau \sim r_{0}^{4} \rightarrow \frac{(N^{1/3}r_{0})^{4}}{N} \sim N^{1/3}r_{0}^{4} \sim N^{1/3} \tau.
\end{eqnarray}
Again, we have a correct dimensional analysis.

Note that, although $\tau/r_{0}$ is invariant under the large $N$ re-scaling, each conformaly equivalent distances should be stretched via $\sqrt{N}$ factor. Therefore, in general, in the $N=1$ limit, the duplication may be observed if one can send a signal between the time
\begin{eqnarray}
\Delta t \sim \exp{-\frac{\tau}{r_{0}}} \sim \exp{-\frac{\tau}{M}},
\end{eqnarray}
where $\tau$ is the information retention time ($\sim M^{3}$); in the large $N$ re-scaled case (Equation~\ref{eq:res}),
\begin{eqnarray}
\Delta t' \sim \sqrt{N} \exp{-\frac{\tau'}{r'_{0}}} \sim \sqrt{N} \exp{-\frac{\tau}{M}}
\end{eqnarray}
is the condition.

From the uncertainty relation, the required energy becomes
\begin{eqnarray}
\Delta E' \sim \frac{1}{\sqrt{N}} \exp{\frac{\tau}{M}},
\end{eqnarray}
and since the consistency of complementarity requires $\Delta E' > M' = \sqrt{N} M$,
the consistency condition becomes
\begin{eqnarray}
\exp{M^{2}} > N M.
\end{eqnarray}
This condition can be violated by assuming a sufficiently large $N \sim \exp M^{2}$ \cite{Yeom:2008qw}.

\subsection{Scrambling time $\tau_{\mathrm{scr}} \sim M \log M$}

The scrambling time is $\sim M \log M \sim M \log S$, where $S$ is the entropy of the black hole \cite{Hayden:2007cs}. Then, in fact, the re-scaling is for $M \log S/\hbar$, and hence, the re-scaling is $\sim \sqrt{N} M \log \sqrt{N}M$. (Therefore, the time is not conformaly invariant.)

Then, in a large $N$ universe, the time scale becomes
\begin{eqnarray}
\Delta t' \sim \sqrt{N} \exp{\left(-\frac{M \log \sqrt{N}M}{M}\right)} \sim \sqrt{N} \exp{\left(-\log \sqrt{N}M\right)}.
\end{eqnarray}
From the uncertainty relation, the required energy becomes
\begin{eqnarray}
\Delta E' \sim \frac{1}{\sqrt{N}} \exp{\log \sqrt{N}M},
\end{eqnarray}
and since the consistency of complementarity requires $\Delta E' > \sqrt{N} M$,
the consistency condition becomes
\begin{eqnarray}
M > \sqrt{N} M.
\end{eqnarray}
Of course, this condition can be violated by assuming a sufficiently large $N$.\footnote{If the scrambling time is $\alpha M \log M$ with a suitable constant $\alpha$, then $\Delta E' > \sqrt{N} M$ may hold. However, the meaning of the scrambling time is a statistical notion, and hence it should not crucially depend on a constant factor $\alpha$. The scrambling may happen after the time $(1/2) M \log M$, and also may happen after the time $2 M \log M$. Therefore, we think that the previous argument is sufficient to violate black hole complementarity.}

In this sense, the black hole complementarity principle can be violated even if we consider a Schwarzschild black hole.

\subsection{Comparison with Dvali's idea}

In some previous papers of Dvali and colleagues \cite{Dvali:2007hz}, they already noticed that large $N$ is harmful to black hole complementarity and hence there should be a relation between the number $N$ and the cutoff scale such as
\begin{eqnarray}
l_{\mathrm{cutoff}} = \sqrt{N} l_{\mathrm{Pl}}.
\end{eqnarray}

To understand their argument, let us think a black hole with large $N$ and mass $M$. Then the information retention time is
\begin{eqnarray}
\tau \sim \frac{M^{3}}{N}
\end{eqnarray}
and the required time $\Delta t$ becomes
\begin{eqnarray}
\Delta t \sim \exp{-\frac{\tau}{M}} \sim \exp{-\frac{M^{2}}{N}}.
\end{eqnarray}
Then, via the uncertainty relation, the required energy $\Delta E$ is
\begin{eqnarray}
\Delta E \sim \exp{\frac{M^{2}}{N}}.
\end{eqnarray}
The consistency condition is $\Delta E > M$ or
\begin{eqnarray}
M^{2} > N \log M.
\end{eqnarray}
Of course, if $M$ is smaller than $\sqrt{N}$ (in the Planck units), then the relation cannot hold. Therefore, Dvali and colleagues concluded that a semi-classical black hole should be greater than $\sqrt{N}$.

In this paper, we argued that large $N$ violates black hole complementarity. However, we used a different context. Dvali thought a black hole $M$ with large $N$, while we thought a black hole $\sqrt{N} M$ with large $N$. Therefore, even though the cutoff is stretched to $\sqrt{N}$, our argument is still valid since we think a sufficiently large black hole than the cutoff scale: $\sqrt{N}M \gg \sqrt{N}$.

\subsection{How large $N$ is required to violate black hole complementarity?}

Note that, in a Schwarzschild black hole of $4$-dimensions, we can suggest two meaningful time scales:
the information retention time $\tau \sim M^{3}$ and the scrambling time $\tau \sim M \log M$.
Here, the required $N$ to violate black hole complementarity is on the order of $\exp \tau/M$.
Therefore, in terms of the information retention time, $\exp M^{2}$ fields are required\footnote{It is meaningful to compare the same factor for charged black holes \cite{Hong:2008mw}.};
whereas in terms of the scrambling time, $\sim 1$ fields are required.
This is not so strange since the scrambling time was regarded as the marginal time to violate black hole complementarity.


One question is whether our real universe violates black hole complementarity or not. According to the scrambling time, it is reasonable to think that it is possible to violate black hole complementarity even in our universe with standard model particles $N \sim 100$.
Or, even though such value is overestimated, string theory will allow a universe which violates black hole complementarity with sufficiently and reasonably enough $N$, in the phenomenologically viable limit \cite{Dvali:2007hz}.

\subsection{Locality bound in the large $N$ limit}

Traditionally, there are mainly three options with respect to the information loss problem:
(1) information is attached by Hawking radiation,
(2) there remains a remnant of very long lifetime, and (3) one cannot regain information from a black hole.
If one does not choose (3), then the remaining possibilities are (1) and (2).
In terms of semi-classical gravity, however, (1) cannot be obtained \cite{Giddings:2001pt}.
Therefore, one needs to violate some assumptions of semi-classical gravity;
the easiest way is to violate locality \cite{Giddings:2001pt}.
However, if one considers the violation of locality only for the area near a singularity, it cannot be helpful to obtain (1).
Therefore, if one wants to obtain (1), we expect that there must be an effect of violation of locality for large black holes, which is apparently semi-classical.
Note that, (1) or violation of locality naturally implies black hole complementarity \cite{Giddings:2001pt};
also, black hole complementarity implies a violation of locality \cite{Lowe:1995ac}.

The violation of locality should be related to a strong gravitational effect.
Giddings tried to quantify this strong gravitational effect \cite{Giddings:2001pt}.
Let us assume that, for example, two particles are generated from a gravitational background.
Each particle has approximately position $(x,y)$ and momentum $(p,q)$ in the center of mass frame.
The suggested locality bound is then
\begin{eqnarray}
|x - y| \gtrsim |p + q|,
\end{eqnarray}
for $4$-dimensions.
If this does not hold, then one may interpret that the gravitational effect is sufficiently strong and violates locality.

Let us assume the creation of two particles with positions $(x,y)$ and momentum $(p,q)$ where the locality bound does not hold.
However, the positions and momentum should be solutions within the background metric, and they could be re-scaled by a large $N$.
Since the locality bound relation is scaled by $\sqrt{N}$ for both the left and right hand sides,
the direction of the inequality is not changed by the re-scaling.
However, we know that the gravitational effect becomes smaller and smaller in the large $N$ limit.
Therefore, a reasonable interpretation is that a certain quantum event that violates locality bound does not occur in the large $N$ limit of a large black hole background.

This conclusion is consistent with previous subsections.
Giddings suggested that the effect of violation of locality will be dominant
by the time of the order of the information retention time $M^{3}$ or the scrambling time $M \log M$ \cite{Giddings:2001pt}.
These times are meaningful for an asymptotic observer, but not meaningful for a free-falling observer.
For example, if two observers can communicate with each other,
then the scrambling near a horizon is meaningless, since the free-falling observer does not be scrambled near the horizon.
Of course, in a small $N$ universe, according to arguments of black hole complementarity, two observers could not communicate.
However, in a large $N$ limit, we know that they can communicate freely.
Therefore, it is a consistent interpretation that, in the large $N$ limit, these time scales are meaningless and the locality bound must not be violated with a large black hole background.

The violation of locality for semi-classical black holes then cannot be a fundamental resolution of the information loss problem.
Of course, it may be helpful in understanding certain small $N$ black holes with certain causal structures;
but it cannot be applied to explain unitarity of large $N$ black holes with the same causal structures.
For a given causal structure, and for all possible $N$, if the violation of locality is essential to explain unitarity,
then the violation should be near the singularity.

\subsection{Is an entanglement helpful to the information loss problem?}

As another approach, we can turn to ideas of quantum information theory.
This may be helpful to address the information loss problem as well as black hole complementarity \cite{Hong:2008ga}.

For example, a proposal of Horowitz and Maldacena \cite{Horowitz:2003he} assumes a final state of the singularity.
They assume that the Hawking radiation is maximally entangled between the in-going and the out-going part.
One potential concern here is that the in-falling matter may destroy the entanglement between the inside and the outside \cite{Gottesman:2003up}.
If this is true, then the proposal will not hold \cite{Lloyd:2004wn}.
However, it remains unclear whether interactions between the in-falling matter and the in-going Hawking radiation can be implemented
to violate the proposal.
If there is a kind of limitation to use the interactions,
then one may state that even if there is a potential problem regarding maximal entanglement, it will work in real situations.

\begin{figure}
\begin{center}
\includegraphics[scale=0.8]{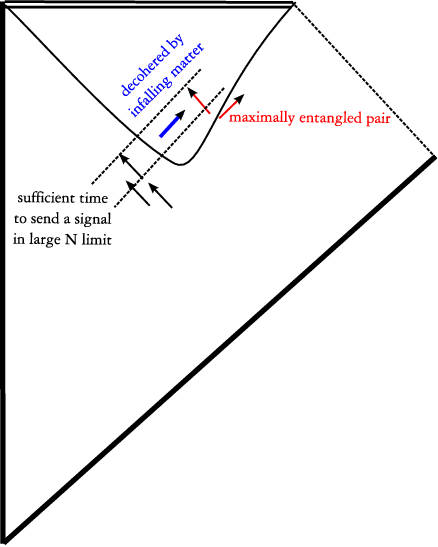}
\caption{\label{fig:entanglement} Arbitrary decoherence can be implemented if there are sufficient $N$.}
\end{center}
\end{figure}

In our re-scaling setup, we use the same Penrose diagram as the number of massless fields grows.
Here, since a length of any two points on the Penrose diagram becomes longer and longer by a factor $\sqrt{N}$,
one can send a signal between arbitrary two points on the Penrose diagram with reasonably small energy.
This implies that we can destroy the entanglement of the proposal as desired (Figure \ref{fig:entanglement}).
Of course, more concrete discussion about this issue is necessary;
nevertheless, a naive expectation is that, if one assumes a large $N$,
one can destroy the entanglement of Hawking radiation,
and this will break the core assumption of the proposal.
Hence, it is unclear whether a quantum information theoretical resolution of the information loss problem
is a fundamental resolution.

\subsection{Singularity and remnant picture}

What will happen if black hole complementarity is not true?

Let us assume that an ideal observer who can control all outcomes of a black hole can reconstruct the original information in principle.
Then, if the area is proportional to the physical entropy of a black hole,
black hole complementarity is inevitable from Page's argument \cite{Page:1993wv};
information should escape around the information retention time, and even in this time, a black hole can be sufficiently large.
Therefore, information should be attached by Hawking radiation,
and there can be two ideal observers, where one is an in-falling observer and the other is an out-going observer.
If the no cloning theorem is correct, then they should not compare observations.
However, as we discussed, in the large $N$ universe, black hole complementarity cannot be true,
and the violation of locality for large black holes cannot be helpful for information conservation.

Then which assumption is invalid in the previous picture? If unitarity is correct, there remains two possibilities.
\begin{enumerate}
  \item There is no such ideal observer who can control all outcomes of a black hole \cite{Hawking:2005kf}.
  \item The area is not the physical entropy but just an apparent entropy;
  therefore, the real entropy should be calculated by the inside degrees of freedom,
  or calculation of real physical entropy is meaningless \cite{Yeom:2008qw}.
\end{enumerate}
Note that, since the first possibility implies that the outside observer cannot reconstruct the information from a black hole,
if it is correct, then the assumption that the physical entropy is proportional to its area becomes meaningless.
Thus, if the first possibility is correct, then the second possibility should be correct, too.
Now, it is inevitable to accept the second possibility.

Therefore, first, let us assume the negation of the first possibility and assume the second possibility;
that is, the real information retention time is not the moment when the initial area of a black hole decreases to its half value.
Hence, information will not be contained by the Hawking radiation, and all information should be contained by the final remnant.
Then the final remnant, which has in general small area, should contain very large entropy;
then, in general, the outcome of a small remnant should have very large entropy,
and the wavelength of the outcome of the final remnant then should be very long
if the outcomes contain all initial information \cite{Giddings:1993km}.
Therefore, it is equivalent to the remnant picture, where the remnant has a very long lifetime.

We will not consider details of this idea; however, one important comment is that, to study this possibility,
it is necessary to solve the problem of singularity.
In black hole complementarity, it is not necessary to consider the trouble of singularity;
however, in the remnant picture, one needs to calculate the entropy of the final remnant,
and full calculations around the singularity are required.
If full calculation around the singularity is possible,
then one can extend the causal structure beyond the singularity.
Some authors have contended that the information loss problem can be resolved
if the causal structure beyond a singularity is solved \cite{Nielsen:2008kd} (we call this the causal structure picture).
This assertion is partly correct, but not entirely; they did not consider a problem of entropy.

In conclusion, we can suggest a very cautious but probable comment on the information loss problem.
One possibility is that there is no ideal observer who can reconstruct the original information;
the other possibility is that information is retained by a long lifetime remnant.
For the latter idea, one needs to study the entropy near a singularity as well as the causal structure beyond the singularity.
As the authors understand, the regular black hole picture (\cite{Nielsen:2008kd} and references therein)
or the causal structure picture \cite{Ashtekar:2005cj} are equivalent with the remnant picture;
the former models will inherit the same problems, if the remnant picture has problems of entropy.

\section{\label{dis}Discussion: Toward singularity}

We claim that large $N$ semi-classical gravity is a useful tool to examine the information loss problem.
We can define re-scaling between an $N=1$ universe and a large $N$ universe.
Here, re-scaling by large $N$ fields can preserve the causal structure of any semi-classical black hole solution in the $N=1$ universe.

If a resolution of the information loss problem must be for the final stage of the black hole, the present paper is not relevant.
However, if the resolution could be applied to even a semi-classical black hole, then our discussion can be meaningful.
Any idea resolving the information loss problem should be valid in a large $N$ setup.
On these grounds, it is possible to test the consistency of black hole complementarity and violation of locality for semi-classical black holes; black hole complementarity and violation of locality do not hold in the large $N$ limit.

Thus, we suspect that the essence of the information loss problem may be located around the singularity at the final stage of the black hole evolution, where a new perspective on the causality is required and hence the semi-classical argument is no longer valid.
Therefore, one sound conclusion is that \textit{the information loss problem will be resolved by using an idea that resolves the problem of the singularity}; in other words, if it does not resolve the singularity, like black hole complementarity,
it cannot be a fundamental resolution to the information loss problem.

\section*{Acknowledgment}
DY would like to thank Ewan Stewart, Seungjoon Hyun, Piljin Yi and Xian-Hui Ge for discussions and encouragements. HZ would like to thank Bayram Tekin. DY and HZ are supported by BK21 and the Korea Research Foundation Grant funded by the Korean government (MOEHRD; KRF-313-2007-C00164, KRF-341-2007-C00010). DY is supported by the National Research Foundation of Korea(NRF) grant funded by the Korea government(MEST) through the Center for Quantum Spacetime(CQUeST) of Sogang University with grant number 2005-0049409. HZ is supported by TUBITAK research fellowship program for foreign
citizens.

\section*{\label{real}Appendix. Realization of large $N$ setup}

In this Appendix, we list possible roots to obtain a large $N$ from string theory or string inspired models.

\subsection*{KKLT}

The KKLT scenario \cite{Kachru:2003aw} is used to compactify $6$-dimensions to a Calabi-Yau manifold,
and some combinations of branes and anti-branes should be assumed.
One may have to stack up D3-branes and then the large number of branes will give large number of fields.

Let us assume that there are $\mathcal{N}$ D3-branes with a weak coupling limit and perpendicular directions are compactified by volume $V_{10-D}$.
Here, $g^{2}=\exp{\langle\phi\rangle}$ is the coupling (we follow the notations of \cite{Gasperini}). We choose the following condition:
\begin{eqnarray}
g^{2} \mathcal{N} \lesssim 1.
\end{eqnarray}
Then, the number of fields will be on the order of $\mathcal{N}^{2}$, since $\mathcal{N}$ D-branes induces $SU(\mathcal{N})$ theory.
According to Dvali, if there are many species of particles, each mass of particles should be on the order of $1/\sqrt{\mathcal{N}^{2}}$ \cite{Dvali:2007hz};
therefore, they are effectively massless.

Also, in this setup, the cosmological constant can be fine-tuned by tuning the number of anti-D3-branes.
Therefore, this scenario is the most natural estimation for our purposes.

One additional requirement is that, to control all orders of quantum effects, one needs to assume the tadpole cancelation condition:
\begin{eqnarray}
\frac{\chi(X)}{24} = N_{D3} + \frac{1}{2 \kappa_{10}^{2} T_{3}} \int_{M}H_{3}\wedge F_{3},
\end{eqnarray}
where $\chi(X)$ is the Euler characteristic of a manifold $X$ and $N_{D3}$ is the number of net D3-branes.
Therefore, there may be a limitation to choose a Calabi-Yau manifold for such a large number of Euler characteristics to cancel out the large number of net D3-branes.

\subsection*{Brane world: Large black holes in weak energy limit}

If $D=5$, and if the fifth dimension is compactified with size $r_{c}$, one can derive the Randall-Sundrum scenario \cite{Randall:1999vf}.
The basic action is
\begin{eqnarray}
S = S_{gravity} + S_{brane} + S_{brane'},
\end{eqnarray}
where
\begin{eqnarray}
S_{gravity} = \int dx^{4} \int dy \sqrt{-G} \{ -\Lambda + 2 M_{5}^{3} R \},
\end{eqnarray}
where $G$ is the determinant of the metric, $\Lambda$ is the cosmological constant, and $M_{5}$ is the Planck mass of 5-dimensions; and
\begin{eqnarray}
S_{brane} = \int dx^{4} \{ V_{brane} + \mathcal{L}_{brane} \}.
\end{eqnarray}
Then, we naturally obtain
\begin{eqnarray}
M_{5}^{3} \propto g^{-2} \propto \frac{1}{G_{5}} \propto \mathcal{N}.
\end{eqnarray}

According to Randall and Sundrum \cite{Randall:1999vf}, they assume the warped metric ansatz
\begin{eqnarray}
ds^{2} = e^{- 2 k |y|} \eta_{\mu \nu} dx^{\mu} dx^{\nu} + dy^{2}
\end{eqnarray}
with the following conditions:
\begin{eqnarray}
V_{brane} = - V_{brane'} = 24 M_{5}^{3} k
\end{eqnarray}
and
\begin{eqnarray}
\Lambda = - 24 M_{5}^{3} k^{2}.
\end{eqnarray}
If one assumes there is a $r_{c} \rightarrow \infty$ limit, we still have a finite Planck mass:
\begin{eqnarray}
M_{Pl}^{2} = \frac{M_{5}^{3}}{k}\left[1-e^{-2kr_{c}\pi}\right] \rightarrow \frac{M_{5}^{3}}{k}.
\end{eqnarray}

In this setup, the bulk is an anti de Sitter space.
However, if one couples the theory with a scalar field with a potential, a de Sitter space can be derived.
We follow the results of a thorough paper of Shiromizu, Maeda, and Sasaki \cite{Shiromizu:1999wj}.
From the $5$-dimensional Einstein equation, one can derive its $4$-dimensional part.
After assuming the metric ansatz
\begin{eqnarray}
ds^{2} = d\chi^{2} + q_{\mu \nu} dx^{\mu}dx^{\nu}
\end{eqnarray}
and the energy momentum tensor
\begin{eqnarray}
T_{\mu \nu} = -\Lambda g_{\mu \nu} + \delta(\chi) (-\lambda q_{\mu \nu} + \tau_{\mu \nu}),
\end{eqnarray}
one can impose $Z_{2}$ symmetry along the $\chi$ direction.
Here, $\lambda$ and $\tau_{\mu \nu}$ are the vacuum energy and the energy momentum tensor of the brane world.

Then, finally, one can derive the $4$-dimensional Einstein equation:
\begin{eqnarray}
^{(4)}G_{\mu \nu} = -\Lambda_{4} g_{\mu \nu} + 8 \pi G_{4} \tau_{\mu \nu} + \frac{1}{M_{5}^{6}} \pi_{\mu \nu} - E_{\mu \nu},
\end{eqnarray}
where
\begin{eqnarray}
\Lambda_{4} &=& \frac{1}{2 M_{5}^{3}} \left( \Lambda + \frac{\lambda^{2}}{6 M_{5}^{3}} \right), \\
G_{4} &=& \frac{\lambda}{48 \pi M_{5}^{6}}, \\
\pi_{\mu \nu} &=& -\frac{1}{4}\tau_{\mu\alpha}\tau^{\alpha}_{\nu} + \frac{1}{12}\tau \tau_{\mu \nu} + \frac{1}{8}q_{\mu\nu}\tau_{\alpha \beta}\tau^{\alpha \beta} - \frac{1}{24}q_{\mu \nu}\tau^{2},
\end{eqnarray}
and $E_{\mu \nu}$ is the $5$-dimensional Weyl tensor.
Note that $\pi_{\mu \nu}$ and the longitudinal part of the Weyl tensor is on the order of $\tau^{2}$; if $\tau \sim R \ll 1$,
one can neglect them ($R$ is the $4$-dimensional curvature).
Also, the transverse part of the Weyl tensor is negligible as long as $rk \gg 1$, where $k = \lambda / (6 M_{5}^{3})$ for a $\Lambda_{4} \simeq 0$ case.
If we can apply two conditions, the following can be obtained:
\begin{eqnarray} \label{Einstein}
^{(4)}G_{\mu \nu} = -\Lambda_{4} g_{\mu \nu} + 8 \pi G_{4} \tau_{\mu \nu}.
\end{eqnarray}

Now, let us check whether the theory allows the conditions of re-scalings or not.
If there is a sufficient number of D3-branes, the theory on the brane will have $SU(\mathcal{N})$,
and the theory contains massless fields on the order of $N \sim \mathcal{N}^{2}$.
Of course, the semi-classical effects (the Hawking radiation) are dominant on branes; thus, Equation (\ref{Einstein}) naturally induces Equation (\ref{Einstein_largeN}) in a semi-classical sense.

If one chooses
\begin{eqnarray}
\lambda \propto \mathcal{N},
\end{eqnarray}
then $k$ is on the order of $1$, and $G_{4}$ is on the order of $1/\mathcal{N} \sim g^{2}$.
Now, one can check the consistency of Equation (\ref{Einstein}). If the gravitational constant is re-scaled in the $\Lambda_{4} \simeq 0$ limit, the following can be obtained (in the Planck units):
\begin{eqnarray}
^{(4)}G_{\mu \nu} = 8 \pi \tau_{\mu \nu} + \pi_{\mu \nu} - E_{\mu \nu}.
\end{eqnarray}
However, after the re-scaling, the terms of order $\tau^{2}$ will be quickly become smaller and smaller on the order of $1/N^{2}$. Also, since $k \sim 1$, if we choose the characteristic size of the black hole to be sufficiently large, $rk \gg 1$ can be easily obtained. Therefore, our setup is consistent with Equation (\ref{Einstein}).

Therefore, in this brane world setup, we can find a correspondence between the semi-classical equation (\ref{semi-classical}) and the brane world setup; if there is a solution from Equation (\ref{semi-classical}), there exists a brane world universe with a sufficiently large $\mathcal{N}$ that the same causal structure is justified.

\subsection*{Brane world: Small black holes}

Now let us look at small black holes in the brane world \cite{Giddings:2001bu} (although the evaporation of a large black hole in an anti de Sitter space is not entirely clear, it will be useful to understand some dimensional analysis on the large $N$ for various dimensions); if we choose a small cosmological constant limit
\begin{eqnarray}
|\Lambda| \ll 1,
\end{eqnarray}
then
\begin{eqnarray}
k^{2} \lesssim \frac{1}{\mathcal{N}}
\end{eqnarray}
holds.
Here, small means that the size of black hole is smaller than the characteristic size ($\sim 1/\sqrt{-\Lambda}$) of the bulk anti de Sitter space.

In small black holes, one cannot assume that the physics is confined in $4$-dimensional gravity. Since one can assume that almost all Hawking radiation emits along the brane modes, we can suggest the following Einstein equation
\begin{eqnarray}
^{(5)} G_{\mu \nu} = 8 \pi ^{(5)}T_{\mu \nu} + 8 \pi \hbar ^{(5)}\langle T_{\mu \nu} \rangle,
\end{eqnarray}
where $^{(5)} T_{\mu \nu}$ is the energy-momentum tensor of $5$-dimensions and $^{(5)} \langle T_{\mu \nu} \rangle = \delta(\chi) \langle T_{\mu \nu} \rangle$ is the re-normalized stress tensor for brane modes. If $X$ is any quantity which does not explicitly depend on $\hbar$ with $[X]=L^{\alpha}$, then $X' = X N^{\alpha/3}$ gives the re-scaled solution of the following equation:
\begin{eqnarray}
^{(5)} G_{\mu \nu} = 8 \pi ^{(5)} T_{\mu \nu} + 8 \pi N \hbar ^{(5)} \langle T_{\mu \nu} \rangle,
\end{eqnarray}
since $^{(5)} G_{\mu \nu}$ has a dimension $L^{-2}$, $^{(5)} T_{\mu \nu}$ has a dimension $L^{-2}$, and $^{(5)} \langle T_{\mu \nu} \rangle$ has a dimension $L^{-5}$.

As a dimensional analysis, this scheme can be extended for arbitrary dimensions; if $D$-dimensions are not compactified and the other dimensions are compactified as in the Randall-Sundrum scenario, we can assume the following:
\begin{eqnarray}
M_{D}^{D-2} \propto g^{-2} \propto \frac{1}{G_{D}} \propto \mathcal{N}.
\end{eqnarray}
Also, we apply the next condition for small black holes:
\begin{eqnarray}
\Lambda \propto - M_{D}^{D-2} k^{D-3} \ll 1.
\end{eqnarray}
Then
\begin{eqnarray}
k \lesssim \mathcal{N}^{-1/(D-3)}
\end{eqnarray}
is reasonable. If combinations of $X$ with $[X]=L^{\alpha}$ is a quantity of the solutions of
\begin{eqnarray}
^{(D)} G_{\mu \nu} = 8 \pi ^{(D)} T_{\mu \nu} + 8 \pi \hbar ^{(D)} \langle T_{\mu \nu} \rangle,
\end{eqnarray}
then
\begin{eqnarray} \label{D-dimension}
X' = X N^{\frac{\alpha}{D-2}},
\end{eqnarray}
are also solutions of a proper $N$ limit.

Therefore, all benefits outlined in Section~\ref{invariance} can be applied to small black holes of the brane world in the large $\mathcal{N}$ limit. Note that Equation~(\ref{D-dimension}) is the correct formula for general dimensions, since our arguments were based on dimensional analysis.

\end{document}